\documentclass[useAMS,usenatbib]{mn2e}

\usepackage{graphicx}
\usepackage{comment}
\usepackage{amsfonts}
\usepackage{amssymb}
\usepackage{amsmath}

\newcommand{\pdiff}[2]{\frac{\partial #1}{\partial #2}}
\newcommand{\diff}[2]{\frac{d #1}{d #2}}
\usepackage{color}

\title[Radiative losses and cut-offs of energetic particles at relativistic shocks]
{Radiative losses and cut-offs of energetic particles at relativistic shocks}
\author[Paul Dempsey and Peter Duffy]{Paul Dempsey\thanks{E-mail:
paul.dempsey@ucd.ie ; peter.duffy@ucd.ie } and Peter
Duffy\footnotemark[1]\\
UCD School of Mathematical Sciences, University College Dublin, Belfield, Dublin 4, Ireland.}
\begin{document}

\date{Accepted 2007 March 28. Received 2007 March 21; in original form 2007 January 24}

\pagerange{\pageref{firstpage}--\pageref{lastpage}} \pubyear{????}

\maketitle

\label{firstpage}

\begin{abstract}
We investigate the acceleration and simultaneous radiative losses of electrons in the vicinity 
of relativistic shocks. Particles undergo pitch angle diffusion, gaining energy as 
they cross the shock by the Fermi mechanism and also emitting synchrotron radiation 
in the ambient magnetic field. A semi-analytic approach is developed which allows us
to consider the behaviour of the shape of the spectral cut-off and the variation of 
that cut-off with the particle pitch angle. The implications for the synchrotron emission
of relativistic jets, such as those in gamma ray burst sources and blazars, are discussed.
\end{abstract}


\section{Introduction}

The role of radiative losses in determining the spectra from non-thermal 
sources has been well understood in the non-relativistic shock limit since the 
work of \citet{WDB1984} and \citet{HeavensMeisenheimer87}. Their 
results were in broad agreement with the natural expectation that there 
would be a cut-off in the spectrum, at the shock, and at a momentum where acceleration and 
loss timescales are equal, with the shape of this cut-off depending critically on the momentum 
dependence of the particle scattering. 
Subsequently, as the particles are advected downstream, and are no longer 
efficiently accelerated by the shock, the spectra steepens at momenta where the particles 
have had sufficient time to cool. At a strong, nonrelativistic shock the differential 
number density of particles at energies where radiative cooling is unimportant is a power law
with $N(E)\propto E^{-0.5}$ with a corresponding intensity of $I_\nu\propto\nu^{-0.5}$ for the 
emitted synchrotron radiation. At higher momenta, where cooling becomes important, the spectrum steepens 
so that the radiation, beyond a break frequency $\nu_b$, is $I_\nu\propto\nu^{-1}$ up to a 
critical frequency, $\nu_{c}$ corresponding to cut-off of the particle spectrum. The position of 
$\nu_b$ depends on position away from the shock; decreasing downstream as the particles have more
time to cool. The {\it observed} emission is therefore dependent on the spatial resolution with which 
the source is observed as discussed in \citet{HeavensMeisenheimer87}. 
 The results in the existing literature refer only to 
non-relativistic flows and are of great use in analysing the spectra from 
supernovae and the jets of some active galaxies (AGN). However, a number 
of objects of astrophysical importance, such as AGN jets, microquasars and 
gamma-ray bursts, contain flows which have bulk relativistic motion and the purpose of this paper 
is to examine the breaks, cut-offs and emission for such sources.

While the first order Fermi process at relativistic shocks contains the same basic physics
as in the nonrelativistic case, i.e. scattering leading to multiple shock crossings competing with a finite 
chance of escape downstream, the anisotropy of the particle distribution complicates the analysis 
considerably (\citet{KS1987}, \citet{HD1988} and \citet{KirkEtAl2000}). The inclusion of self-consistent 
synchrotron losses will, as in the nonrelativistic limit, modify the spectrum at high momenta but we 
would also expect pitch angle effects to become apparent in the position of the cut-off and the 
emission  itself. In order to motivate our treatment of this problem we first
discuss the nonrelativistic shock limit in section 2, including the emission from a spatially integrated 
source. Section 3 then presents the analysis of synchrotron losses at relativistic shocks with particular 
emphasis on the shape of the momentum cut-off. We conclude with a discussion in section 4.

\section{Nonrelativistic Shocks}

The effect of synchrotron losses on the energetic particle distribution in the 
presence of nonrelativistic shocks is demonstrated rigorously in \citet{WDB1984}. 
However a simpler approach is described in \citet{HeavensMeisenheimer87} provided 
synchrotron losses are not considered important at the injection energies. 
We will follow this approach here, although we shall introduce 
a slightly different definition of the cut-off momentum.

In the presense of a magnetic field charged particles emit synchrotron 
radiation with an energy loss rate given by
\begin{align}
 \diff{p}{t}=-a_s B^2p^2=-\lambda p^2
\end{align}
where $a_s$ is a positive constant. The radiative loss timescale is therefore
$t_{\rm loss}= 1/(\lambda p)$. In the steady state, and in the presence of a  
nonrelativistic flow $u$, energetic particles obey 
a transport equation describing advection, diffusion, adiabatic compression and radiative losses,
\begin{align}
\label{transDiffusionApproxSynch}
u\pdiff{f}{z} - \pdiff{}{z}\left(\kappa\pdiff{f}{z}\right) {-\frac{1}{3}p\pdiff{u}{z}\pdiff{f}{p} }-  \frac{1}{p^2}\pdiff{}{p}\left(\lambda p^4 f\right) =0.
\end{align}

In the presence of a nonrelativistic shock front where the upstream flow speed is $u_-$ and that 
downstream is $u_+$ the acceleration timescale is
\begin{equation}
t_{\rm acc}=\frac{3}{u_--u_+}\left(\frac{\kappa_-}{u_-} + \frac{\kappa_+}{u_+}\right).
\end{equation}
At momenta for which $t_{\rm acc}\ll t_{\rm loss}$ the phase space density will be a 
simple power law with $f\propto p^{-s}$ where $s=3u_-/(u_--u_+)$.

\subsection{Momentum Cut-off}

The spectrum will steepen at momentum $p^*$ where $t_{\rm acc}(p^*)=t_{\rm loss}(p^*)$. In the case of 
momentum independent diffusion this gives
\begin{equation}
p^*=\frac{u_--u_+}{3}\left({\frac{\lambda_-\kappa_-}{u_-}}+
\frac{\lambda_+\kappa_+}{u_+}\right)^{-1}.
\end{equation}

In the case of a relativistic shock this result no longer strictly holds since the 
acceleration timescale defined above is only valid for nonrelativistic flows. Nevertheless, we will 
use this definition of $p^*$ throughout the paper for the sake of comparison.

However,  we require a general definition of the cut-off momentum that can be applied in the relativistic 
limit. An obvious alternative is to define the momentum at which the local spectral index, 
$\partial\ln f/\partial\ln p$, becomes $s+1$ but, as we shall see, it is necessary to perform a 
Laplace transform of the transport equation to proceed with this problem and it is more straightforward
to define the cut-off in terms of spectral steepening of the Laplace transformed spectrum.
In order to motivate such a definition we solve the nonrelativistic shock 
acceleration problem in the presence of synchrotron losses by first making the substitutions
$W\equiv p^4f$ and $y\equiv 1/p$ so that the transport equation, either upstream or downstream of the shock
where adiabatic losses are zero, becomes
\begin{align}
u\pdiff{W}{z} - \pdiff{}{z}\left(\kappa\pdiff{W}{z}\right) + \pdiff{}{y}\left(\lambda W\right)=0.
\end{align}
Taking the Laplace transform with respect to $y$
\begin{align}
 {\hat W}(k,z)= \int_{0}^{\infty}W(y,z)\exp(-y k) dy
\end{align}
and using the fact that losses prevent any particles achieving infinite energy,
 i.e. $W(0,z)=0$, the transformed transport equation is
\begin{align}
u\pdiff{{\hat W}}{z} - \pdiff{}{z}\left(\kappa\pdiff{{\hat W}}{z}\right) + \lambda k{\hat W}=0.
\end{align}
in the case of a momentum independent diffusion coefficient. Since the distribution function must be
bounded infinitely far upstream and downstream, the solution becomes
\begin{align}
 {\hat W}_\pm=A_\pm(k)\exp\left(\frac{1{\mp} \sqrt{1+\omega_\pm k}}{2}\frac{u_\pm}{\kappa_\pm} z\right)\label{Wup}
\end{align}
where we have introduced
\begin{align}
\omega_\pm \equiv \frac{4\lambda_\pm\kappa\pm}{u_\pm^2}.
\end{align}

The isotropic and anisotropic parts of the particle distribution function must match up at the shock {giving},
\begin{align}
 f_-(p,0)&=f_+(p,0)\label{isomatch}\\
\kappa_-\frac{\partial f_-}{\partial z}  + \frac{u_-}{3}p\frac{\partial f_-}{\partial p}&= \kappa_+\frac{\partial f_+}{\partial z}  + \frac{u_+}{3}p\frac{\partial f_+}{\partial p}\label{fluxcont}.
\end{align}
Multiplying the isotropic boundary condition by $p^4$, making the 
substitutions as above and taking the Laplace transform with respect to $1/p$ gives 
\begin{align}
 {\hat W}_-(k,z=0)={\hat W}_+(k,z=0)
\end{align}
which in turn gives
\begin{align}
\label{ApmMatch}
 A_-(k)=A_+(k)\equiv A(k).
\end{align}
The {flux continuity condition (\ref{fluxcont})} becomes
\begin{align}
\label{laplaceAk}
 A(k)=A_0k^{3-s}\exp\left(-s\sqrt{1+\omega_-k}-(s-3)\sqrt{1+\omega_+k}\right)\times\nonumber\\ 
\left(1+\sqrt{1+\omega_-k}\right)^{s}\left(1+\sqrt{1+\omega_+k}\right)^{s-3}.
\end{align}

In the absence of synchrotron losses, $\omega_\pm=0$, we have ${\hat W}_\pm (k)\propto k^{3-s}$ which, upon 
inversion, gives $f(p)\propto p^{-s}$ as expected. We can therefore define a function, ${\hat Q}$, by 
${\hat W}=k^{3-s}{\hat Q}$. Recalling that $k$ is the Laplace transformed variable of {\it inverse} momentum 
we define the cut-off momentum, $p_{\rm cut}$, to occur at the point where
\begin{align}
 \left.\pdiff{\ln {\hat Q}}{\ln k}\right\vert_{k=p_{\rm cut}} = -1.
\end{align}
As an illustrative example, consider a power law distribution with a sharp maximum momentum, 
$f(p)\propto p^{-s}H(p_{\rm max}-p)$ with $H$ the Heaviside function. In this case we have
\begin{align}
 W=y^{s-4}H\left(y-\frac{1}{p_{\rm max}}\right)
\;\;\;\Rightarrow\;\;\;
 {\hat W} = e^{-\frac{k}{p_{\rm max}}} \Gamma(s-3) k^{3-s}
\end{align}
where $\Gamma$ is the Gamma function. With ${\hat Q}\propto \exp(-k/p_{\rm max})$ we then 
have $p_{\rm cut}=p_{\rm max}$ as required physically in this simple case.

\begin{figure}
 \centering
	\includegraphics[width=.9\columnwidth]{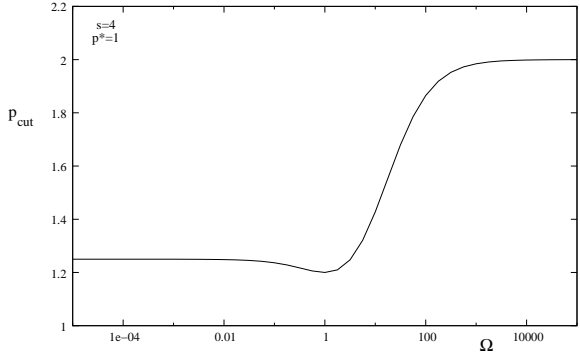}
	\caption[Cut-Off Momentum]{The cut-off momentum, $p_{\rm cut}$ as a 
function of $\Omega$ for fixed equilibrium momentum, $p^\ast=1$, 
and spectral index, $s=4$.}
	\label{fig: pcuts4pe1}
\end{figure}

Returning to the solution of the shock problem we have from equation \ref{laplaceAk}
\begin{align}
 \left.{\hat Q}\right\vert_{z=0}=\exp\left(-s\sqrt{1+\omega_-k}-(s-3)\sqrt{1+\omega_+k}\right)\nonumber\\ \times\left(1+\sqrt{1+\omega_-k}\right)^{s}\left(1+\sqrt{1+\omega_+k}\right)^{s-3}.
\end{align}
Defining
\begin{align}
\Omega&=\omega_+/\omega_-\\
\chi&=\sqrt {\left( {s}^{2}+2\,\,{s}^{2}{\Omega}+{{\Omega}}^
{2}{s}^{2}-6\,{{\Omega}}^{2}s+2\,{\Omega}\,s+9\,{{\Omega}}^{2}
-8\,{\Omega}\, \right)}\nonumber\\&\qquad\times  \left( s-3 \right)
\end{align}
gives
\begin{align}
 p_{\rm cut}=\frac{2}{\omega_-}\Bigg(&\frac{s^4(2+2\Omega)-s^3(5+11\Omega)+s^2(5+8\Omega-2\chi)}{\left(s^2(1-\Omega)+6\Omega s - 9\Omega\right)^2}\nonumber\\&+
\frac{s(33\Omega+\chi)-36\Omega}{\left(s^2(1-\Omega)+6\Omega s - 9\Omega\right)^2}\Bigg).
\end{align}
This is always greater than $p^*$,
\begin{align}
 p^\ast = \frac{4}{\omega_-} \frac{1}{s+(s-3)\Omega}
\end{align}
as can be seen from figure \ref{fig: pcuts4pe1}. The minimum value for $p_{\rm cut}$ occurs for
$\Omega=1$,  and is given by $p_{\rm cut}=\frac{2(s-1)}{2s-3}p^\ast$.

\begin{figure}
 \centering
	\includegraphics[width=.9\columnwidth]{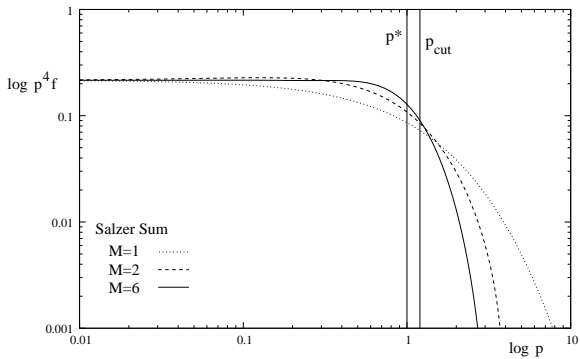}
	\caption[Laplace Inversion Approximation]{Laplace inversion for $s=4$ 
and $\Omega=1$. Using just $M=6$ in the Salzer summation the inversion has 
already converged. The first approximation $M=1$ is exactly the Laplace 
function $\hat{W}k$. We can see how fast the Salzer summation Post-Widder 
inversion converges as $M=2$ is a very reasonably approximation to the 
actual solution.}
	\label{fig: approxsSynch}
\end{figure}

The Laplace inversion {(see Appendix for details)} for $s=4$ and $\Omega=1$ is shown 
in figure \ref{fig: approxsSynch}. Using just $M=6$ in the Salzer summation the 
inversion has already converged. The first approximation $M=1$ is exactly 
the Laplace function $\hat{W}k$. We can see how fast the Salzer summation 
Post-Widder inversion converges as $M=2$ is a good approximation 
to the actual solution.

\begin{figure}
 \centering
	\includegraphics[width=.9\columnwidth]{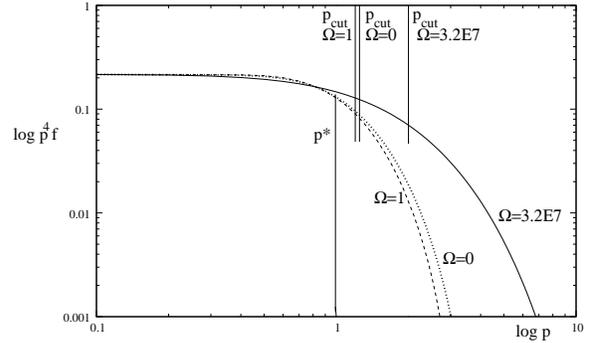}
	\caption[Particle Distributions of Various $\Omega$]{Particle 
Distributions of Various $\Omega$}
	\label{fig: synchSpectraNonRels4}
\end{figure}

Figure~\ref{fig: synchSpectraNonRels4} shows how the particle distribution 
varies with $\Omega$;  if $\Omega\neq1$ the cut-off is broader. While $p^\ast$ 
is independent of $\Omega$, $p_{\rm cut}$ 
increases as the distribution broadens.

The data in figure~\ref{fig: synchSpectraNonRels4} can be fitted by an exponential 
tail to the distribution of the form
\begin{align}
 \exp\left(-\left({p}/{p_{\rm cut}}\right)^{\beta}\right)
\end{align}
where $\beta\sim 2$. Table~\ref{table: paramexploss} shows 
how $\beta$ varies with $\Omega$ for a shock of natural spectral index $s=4$, with  
$\beta$ attaining its maximum value of $2.25$, i.e. the cut-off is sharpest, when $\Omega=1$. 
When $\Omega\gg1$ particles can diffuse in the upstream without losing any energy, allowing a 
greater spread in momentum above $p_{\rm cut}$. 
\begin{table}
 \centering
 \begin{tabular}{|l|l|l|l|}
  \hline
$\Omega$ & $p^\ast$ & $p_{\rm cut}$ & $\beta$ \\
\hline
0 & 1 & 1.25 & 2.25 \\
1 & 1 & 1 & 2.25 \\
9 & 1 & 1.4 & 2 \\
16 & 1 & 1.53 & 1.8 \\
25 & 1 & 1.63 & 1.75\\
$\infty$ & 1 & 2 & 1.5 \\
\hline
 \end{tabular} 
\caption[Parameters for fitting particle spectra]{Parameters for fitting
 particle spectra}
\label{table: paramexploss}
\end{table}

\begin{figure}
\centering
 	\includegraphics[width=.9\columnwidth]{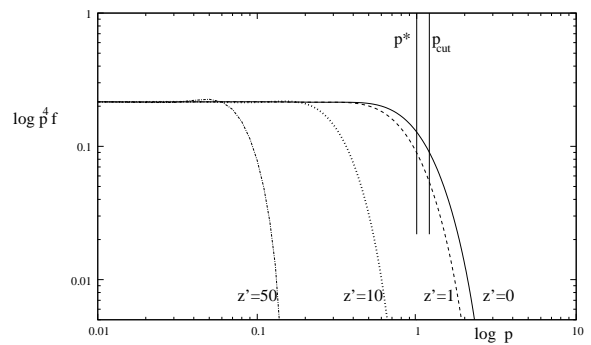}
	\caption[Spatial Effects of Synchrotron Losses]{
The spatial variation of the particle distribution for $\Omega=1$. $z'=\frac{u_+}{\kappa_+}z$.}
\label{fig: spaceDists4Omega1}
\end{figure}
\begin{figure}
\centering
	\includegraphics[width=.9\columnwidth]{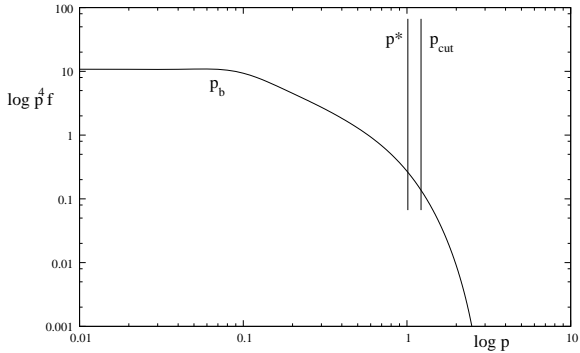}
	\caption[Spatial Effects of Synchrotron Losses -- Integrated Distribution]{
The spatially integrated particle distribution for $\Omega=1$ from $z'=0$ to 
$z'=50$. Note that as well as the particle cut-off there is a spectral break 
earlier were they spectrum softens from $p^{-4}$ to $p^{-5}$. This is 
analogous to synchrotron ageing.}
\label{fig: spaceINTDistz50s4Omega1}
\end{figure}

\subsection{The Integrated Distribution Function and Synchrotron Spectra}

When the source cannot be fully resolved observationally, we must include the contribution
from all particles within some distance $z'$ of the shock in calculating the spatially integrated 
emission. In the case of steady emission from a jet pointing 
towards us, or a completely unresolved source, $z'$ is essentially the source size
in the optically thin limit. For simplicity we assume that the magnetic field downstream of the shock is constant 
although the model can be generalised for more complex cases.

The integrated Laplace distribution function is
\begin{align}
 \hat{T}&=\int_0^{z'}\hat{W} \;dz\nonumber\\
&= \frac{A_0\kappa_+}{u_+} \frac{k^{3-s}\left.{\hat Q}\right\vert_{z=0}
\left(1+\sqrt{1+\omega_+k}\right)}{\omega_+k}\nonumber\\
&\qquad\times\left(1-\exp\left(\frac{1-\sqrt{1+\omega_+k}}{2}\frac{u_+}{\kappa_+}z'\right)\right).
\end{align}
When $z'$ is very small the result is $k^{3-s}$ with a cut-off at high $k$ as 
expected. As $z'$ tends to infinity at low $k$ we have $k^{2-s}$ so the 
spectrum is steepened , with the same high $k$ 
cut-off. For finite values of $z'$ the spectrum starts as $k^{3-s}$ before 
turning into $k^{2-s}$ and finally cutting off. We shall see later that this 
result also holds in real momentum space. 

Figure~\ref{fig: spaceDists4Omega1} shows how the cut-off tends to
lower momenta as we go further downstream. However what is most often 
observed a result of the integrated distribution is shown in figure~\ref{fig: spaceINTDistz50s4Omega1}. 
While the cut-off momentum is the same as at the shock, the distribution 
changes from an initial $p^{-4}$ to a $p^{-5}$ spectrum at some critical 
momentum, $p_{\rm b}$, which depends on $z'$. Here we will consider only synchrotron emission from an ordered magnetic field 
(parallel to the flow). Let $w = \frac{4\pi \nu m_e^3c^2}{3q B}$ where $\nu$ 
is the frequency, $q$ is the charge on the electron, $m_e$ is the electron 
mass and $B$ is the magnetic field strength.
Then given a spatially integrated particle distribution $f\propto p^{-s}
g(p,\mu)$ the total power emitted  per unit frequency is \citep{RPiA}
\begin{align}
 P_{tot}(\omega) \propto \sqrt{1-\mu^2} &\int_0^\infty 
\left(\frac{w}{x\sqrt{1-\mu^2}}\right)^{(s-5)/2} \nonumber\\
&g \left(\sqrt{\frac{w}{x\sqrt{1-\mu^2}}},\mu\right)F(x)\;dx
\end{align}
where $F$ is the first synchrotron function
\begin{align}
 F(x) \equiv x \int_x^\infty K_{\frac{5}{3}}(y)\;dy.
\end{align}

In the case of non-relativistic diffusive shock acceleration $f$ 
downstream of a shock is assumed to be isotropic in which case $g$ is 
independent of $\mu$.

\begin{figure}
 \centering
	\includegraphics[width=.9\columnwidth]{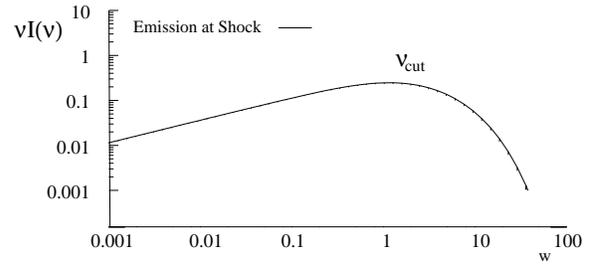}
	\caption[Synchrotron Spectrum at the Shock]{A telescope with very high 
resolution may be able to observe synchrotron radiation at the shock. The only 
spectral feature here is the cut-off at $\nu_{\rm cut}$. Before 
$\nu_{\rm cut}$ the spectrum has shape $I(\nu)\sim \nu^{-1/2}$. 
For this result $\Omega=1$.}
	\label{fig: shockSpectrum}
\end{figure}

\begin{figure}
 \centering
	\includegraphics[width=.9\columnwidth]{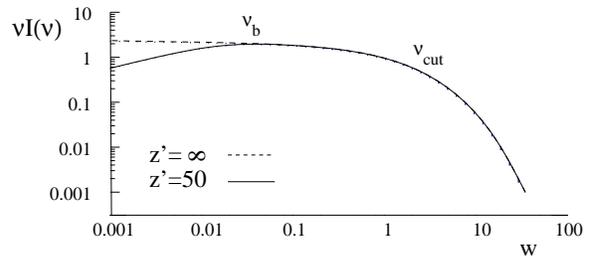}
	\caption[Integrated Synchrotron Spectrum]{Spatially integrated emission 
for $\Omega=1$. The black dashed curve shows the synchrotron spectrum of an 
unresolved object containing a strong shock. The only feature is the 
spectral cut-off at $\nu_{\rm cut}$. Before  $\nu_{\rm cut}$ the spectrum 
has shape $I(\nu)\sim \nu^{-1}$. The solid curve illustrates the 
synchrotron spectrum of a partially resolved source with emission from 
the shock ($z'=0$) to some downstream distance ($z'=50$).}
	\label{fig: integrateSpectrum}
\end{figure}

Figure~\ref{fig: shockSpectrum} shows the emission at the shock. The only feature here is the cut-off hump, which 
is of course related to the particle cut-off, before which $I_\nu\sim\nu^{-(s-3)/2}$. Figure~\ref{fig: integrateSpectrum} 
shows a more realistic plot, that of emission from an extended region. If the region is  infinite in size then the cut-off 
remains the sole feature but the spectrum before the cut-off is different $I_\nu\sim\nu^{-(s-2)/2}$. If the region has finite 
size then a second feature, the spectral break $\nu_b$, appears. Before the break the spectrum goes as $I_\nu\sim\nu^{-(s-3)/2}$ 
while after it it is $I_\nu\sim\nu^{-(s-2)/2}$. Again this is related to the momentum break $p_b$ we see in the particle 
distribution in figure~\ref{fig: spaceINTDistz50s4Omega1}.

\section{Relativistic Shock Acceleration with Losses}

In the case of a relativistic shock, the particle transport equation describing 
advection, pitch angle diffusion and losses becomes
{\begin{align}
\label{basicTransEqn}
\Gamma\, (u+\mu)\frac{\partial f }{\partial z} = \frac{\partial}{\partial \mu}\left(D_{\mu\mu}\frac{\partial f}{\partial\mu}\right) 
+ \lambda g(\mu)\frac{1}{p^2}\frac{\partial (p^4 f) }{\partial p} 
\end{align}
where $\mu$ is the cosine of the pitch angle of the particle and the flow velocity is constant upstream and downstream of the shock. $\lambda = \frac{2\sigma_{T}}{m^2c^2}U_B$ and 
$g(\mu) = 1-\mu^2$ for synchrotron losses in an ordered magnetic field, $\lambda=4\sigma_{\rm T} U_{\rm B}/3$ 
and $g(\mu) = 1$ for synchrotron losses in a tangled  magnetic field, or $\lambda=4\sigma_{\rm T} U_{\rm rad}/3$ 
and $g(\mu) = 1$ for inverse Compton losses.}
Equation (\ref{basicTransEqn}) holds separately upstream and downstream with 
the conditions that the distribution is isotropic infinity far downstream, 
there are no particles infinitely far upstream and the distribution is 
continuous at the shock.
Although we will derive equations for general momentum independent pitch-angle 
diffusion and an arbitrary magnetic field alignment, the figures and results 
produced throughout the rest of this paper are for isotropic diffusion 
$D_{\mu\mu}=D(1-\mu^2)$ in an ordered (longitudinal) magnetic field with 
$\lambda/D=0.1$.

Guided by the treatment of the nonrelativistic case we set $W=p^4f$ and $y=1/p$ so that 
\begin{align}
\Gamma(u+\mu)\pdiff{W}{z} =& \pdiff{}{\mu}\left[D(\mu)(1-\mu^2)
\pdiff{W}{\mu}\right]
-  {\lambda } g(\mu)\pdiff{W}{y}.
\end{align}
Taking the Laplace Transform with respect to $y$ and assuming $W(0,\mu,z)=0$ 
\begin{align}
\Gamma(u+\mu)\pdiff{{\hat W}}{z} =& \pdiff{}{\mu}\left[D(\mu)(1-\mu^2)\pdiff{{\hat W}}{\mu}\right]
-
  {\lambda } g(\mu){k {\hat W}}.
\end{align}
With the spatial and pitch angle variables separable we look for solutions of the form
\begin{align}
{\hat W}(k,\mu,z)=\sum_{i}a_i(k)X_i(k,z)Q_i(k,\mu)
\end{align}
putting this back into the reduced transport equation we get
\begin{align}
\Gamma(u+\mu)\pdiff{X_i}{z}Q_i=(\overline{\mathcal{D}}Q_i)X_i
\end{align}
where we have defined the differential operator $\overline{\mathcal{D}}$ via
\begin{align}
\overline{\mathcal{D}}\Phi = \pdiff{}{\mu}\left[D(\mu)(1-\mu^2)\pdiff{\Phi}{\mu}\right]-  \lambda g(\mu)k\Phi.
\end{align}
Separating $X$ and $Q$ we get the usual
\begin{align}
\Gamma \frac{1}{X_i}\pdiff{X_i}{z}= \Lambda_i(k) = \frac{1}{Q_i(u+\mu)}\overline{\mathcal{D}}Q_i
\\*
\Rightarrow X_i(k,z)= \exp\left(\frac{\Lambda_i(k) z }{\Gamma}\right)
\end{align}
and we have an equation for $Q(k,\mu)$
\begin{align}{
\overline{\mathcal{D}}Q_i - \Lambda_i(k) Q_i(u+\mu)=0.
}
\end{align}
Expanding out the differential operator we get
\begin{align}
\pdiff{}{\mu}\left(D(\mu)(1-\mu^2)\pdiff{Q_i}{\mu}\right)-\left(\Lambda_i(u+\mu)+k\lambda g(\mu)\right)Q_i = 0  
\end{align}
which has regular singularities at $\mu =\pm1$ and so it should be possible 
to find solutions for $Q_i$ on $[-1,1]$ for all $k\in\mathbb{C}$. 

\subsection{Determining the Eigenfunctions}

We know that along the real axis, {$k=x\in\mathbb{R}$}, each $Q_i$ satisfies
\begin{align}
\pdiff{}{\mu}\left(D(\mu)(1-\mu^2)\pdiff{Q_i }{\mu}\right)
-\left(\Lambda_i(u+\mu)+x\lambda g(\mu)\right)Q_i = 0.\label{SLeqnQ}
\end{align}
We define an inner product by:{
\begin{align}
\langle \zeta,\xi\rangle = \int (u+\mu) {\zeta}^* \xi d\mu.
\end{align}}
It can be shown that the $Q_i(x)$ are orthogonal and either real or purely imaginary, and the $\Lambda_i(x)$ are real and distinct.
We can normalise the eigenfunctions such that
 \begin{align}
\langle Q_i, Q_j \rangle = \delta_{i,j}
\end{align}
or considering them as real
\begin{align}
\langle Q_i, Q_j \rangle = \delta_{i,j}{\left(1/2-i\right)/\left|{1}/{2}-i\right|} \equiv \eta_{i,j}.
\end{align}
Then we have (see Appendix for details)
\begin{align}
\pdiff{Q_i }{x} = \lambda \sum_{j\not= i}\frac{1}{\Lambda_j - \Lambda_i}
\left(\int g(\mu) Q_i {Q_j} d\mu\right) Q_j\eta_{j,j}\label{Efunctderiv}
\end{align}
and
\begin{align}
\diff{\Lambda_i}{x} = -\lambda\left(\int g(\mu) Q_i {Q_i} d\mu\right)\eta_{i,i}.\label{Evaluederiv}
\end{align}
{We solve equation \ref{SLeqnQ} at $x=0$ using the Pr{\"u}fer transformation as in \citet{KirkEtAl2000}. We then use 
equations \ref{Efunctderiv} and \ref{Evaluederiv} to find $Q_i(x,\mu)$ and $\Lambda_i(x)$ for $x>0$ using Runge-Kutta methods. }

\begin{figure}
 \centering
	\includegraphics[width=.9\columnwidth]{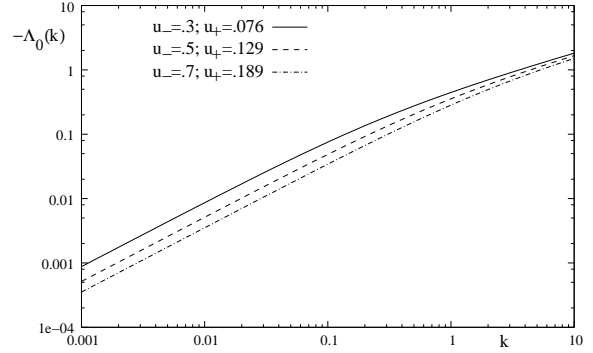}
	\caption[$\Lambda_0(k_+)$ for shock speeds $u_-=.3,.5,.7$]{The zeroth 
order downstream eigenvalue for shock speeds $u_-=.3,.5,.7$. Along the 
$x$-axis we have plotted the logarithm of $k_+$ while along the 
$y$-axis we have the logarithm of $-\Lambda_0(k_+)$. When $k_+=0$ we 
have $\Lambda_0(0)=0$.}
	\label{fig: lambda0}
\end{figure}

\begin{figure}
 \centering
	\includegraphics[width=.9\columnwidth]{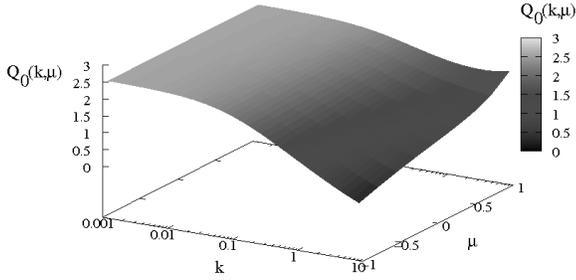}
	\caption[$Q_0(k_+,\mu_+)$ for a shock speed of $.3$]{The zeroth 
order downstream eigenfunction for shock speed $u_-=.3$. Along the $x$-axis 
we have plotted the logarithm of $k_+$ while along the $y$-axis we have 
$\mu_+$. Up the $z$-axis we have plotted $Q_0(k_+,\mu_+)$, which also 
defines the grayscale. Note the anisotropy increases with $k_+$.}
	\label{fig: Qfunct0u3}
\end{figure}

\begin{figure}
 \centering
	\includegraphics[width=.9\columnwidth]{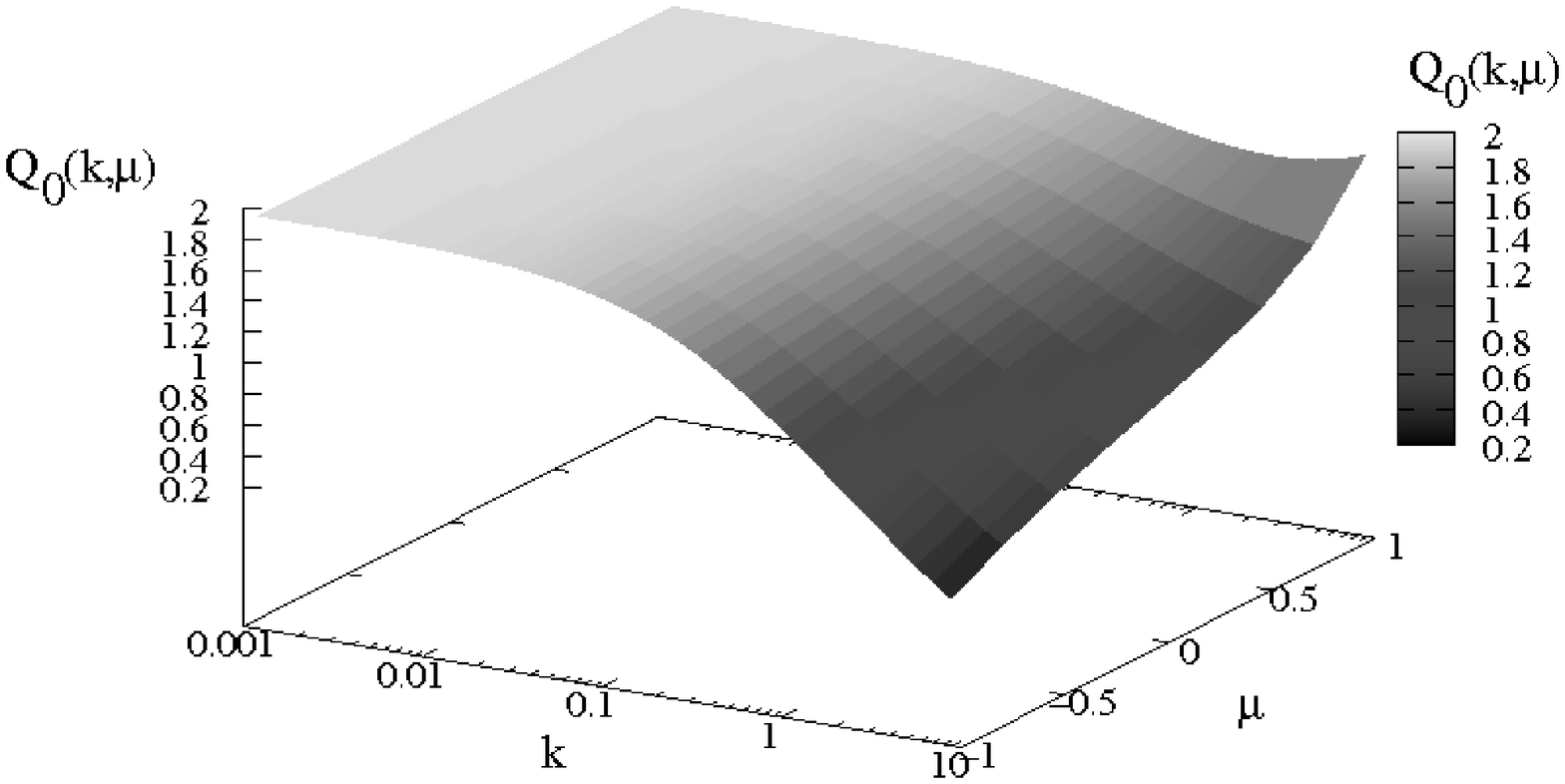}
	\caption[$Q_0(k_+,\mu_+)$ for a shock speed of $.5$]{The zeroth 
order downstream eigenfunction for shock speed $u_-=.5$. Along the $x$-axis 
we have plotted the logarithm of $k_+$ while along the $y$-axis we have 
$\mu_+$. Up the $z$-axis we have plotted $Q_0(k_+,\mu_+)$, which also 
defines the grayscale. Note the anisotropy increases with $k_+$.}
	\label{fig: Qfunct0u5}
\end{figure}

\begin{figure}
 \centering
	\includegraphics[width=.9\columnwidth]{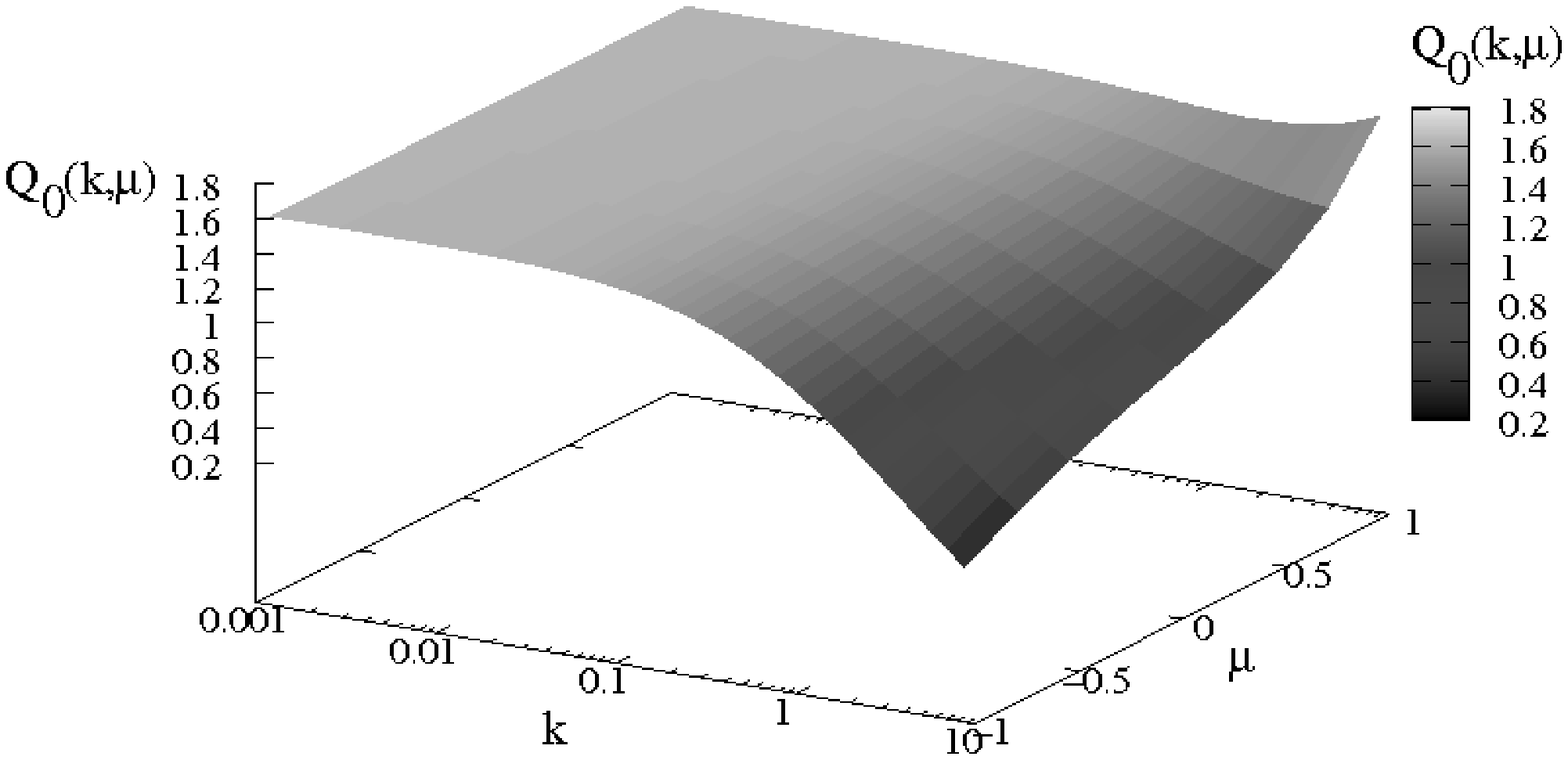}
	\caption[$Q_0(k_+,\mu_+)$ for a shock speed of $.7$]{The zeroth 
order downstream eigenfunction for shock speed $u_-=.7$. Along the $x$-axis 
we have plotted the logarithm of $k_+$ while along the $y$-axis we have 
$\mu_+$. Up the $z$-axis we have plotted $Q_0(k_+,\mu_+)$, which also 
defines the greyscale. Note the anisotropy increases with $k_+$.}
	\label{fig: Qfunct0u7}
\end{figure}

Figures \ref{fig: lambda0}, \ref{fig: Qfunct0u3}, \ref{fig: Qfunct0u5} and 
\ref{fig: Qfunct0u7} show the zeroth downstream eigenvalues and eigenfunctions 
for shocks speeds of .3, .5 and .7. This eigenfunction is the dominant 
component in the downstream distribution function at the shock of such 
mildly relativistic shocks, where we are close to isotropy. Further 
downstream, where the contribution of higher eigenfunctions are more strongly 
damped, so the anisotropy for some $z>0$ is essentially 
that of the zeroth eigenfunction. 
While $\Lambda_0(k_+)$ is initially zero, note from figure \ref{fig: lambda0} 
that it decreases linearly until a certain point which, as we will see 
later, is close to the cut-off momentum. This will play a major role in 
the integrated distribution function and emission. 
Figures \ref{fig: Qfunct0u3}, \ref{fig: Qfunct0u5} and \ref{fig: Qfunct0u7} 
show how anisotropy arises in the zeroth order eigenfunction which is 
isotropic for $k=0$. Given that $k$ is related to the momentum these 
figures show that, since this is the dominant eigenfunction, the 
anisotropy will increase with increasing energy.

\subsection{Shock matching conditions}
Starting from 
\begin{align}
{\hat W}(k,\mu,z)&=\sum_{i}a_{i}(k)X_i(k,z)Q_{i}(k,\mu)\\
&= \sum_{i}a_{i}(k)\exp\left(\frac{\Lambda_i(k) z}{\Gamma}\right)Q_{i}(k,\mu))
\end{align}
{we note that upstream $(z<0)$ we have $ a_{i}=0$ for all $i$ such that $\Lambda_i \leq 0$  and that downstream $(z>0)$ we have  $ a_{i}=0$ for all $i$ such that $\Lambda_i > 0$.}
The distribution function is continuous at the shock,
\begin{align}
f_-(y_-,\mu_-,0)=f_+(y_+,\mu_+,0)
\end{align}
with $(y_-,\mu_-)$ related to $(y_+,\mu_+)$ by a Lorentz transformation of velocity
$u_{\rm rel}=(u_--u_+)/(1-u_-u_+)$,
\begin{align}
 y_- &= \Gamma_{\rm rel}y_+(1+u_{\rm rel}\mu_-).
\end{align}
In terms of $W$ the matching condition becomes
\begin{align}
\Gamma_{\rm rel}^4(1+u_{\rm rel}\mu_-)^4W_-(y_-,\mu_-,0)&=W_+(y_+,\mu_+,0)\label{m28}
\end{align}
and we now need to express this in terms of ${\hat W}$, the Laplace transform with respect to 
$y$. Taking $k_-y_-=k_+y_+$, multiplying the matching condition for $W$ by $\exp(-k_+y_+)$ and 
integrating over $y_+$ gives
\begin{align}
\Gamma_{\rm rel}^3(1+u_{\rm rel}\mu_-)^3{\hat W}_-(k_-,\mu_-,0) =  {\hat W}_+(k_+,\mu_+,0).
\end{align}

\begin{figure}
 \centering
	\includegraphics[width=.9\columnwidth]{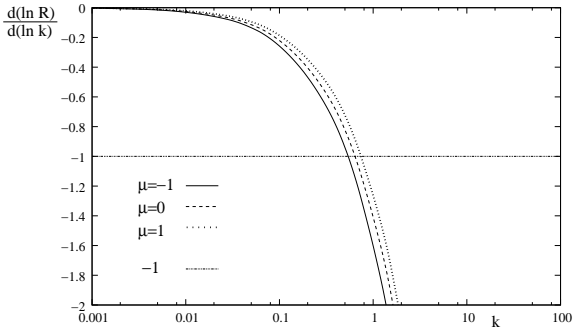}
	\caption[Calculating the cut-off momentum downstream. ($u_-=.3$)]
{Plotted along the $x$-axis we have the logarithm of momentum $k_+$ while 
along the $y$-axis we have $d(\ln R_+)/d(\ln k_+)$ for $u_-=.3$ and 
$R_+=\sum_i b^+_i(k_+)Q_i^+(k_+,\mu_+)$. Note the cut-off depends on $\mu_+$}
	\label{fig: cutoffu_3}
\end{figure}

\begin{figure}
 \centering
	\includegraphics[width=.9\columnwidth]{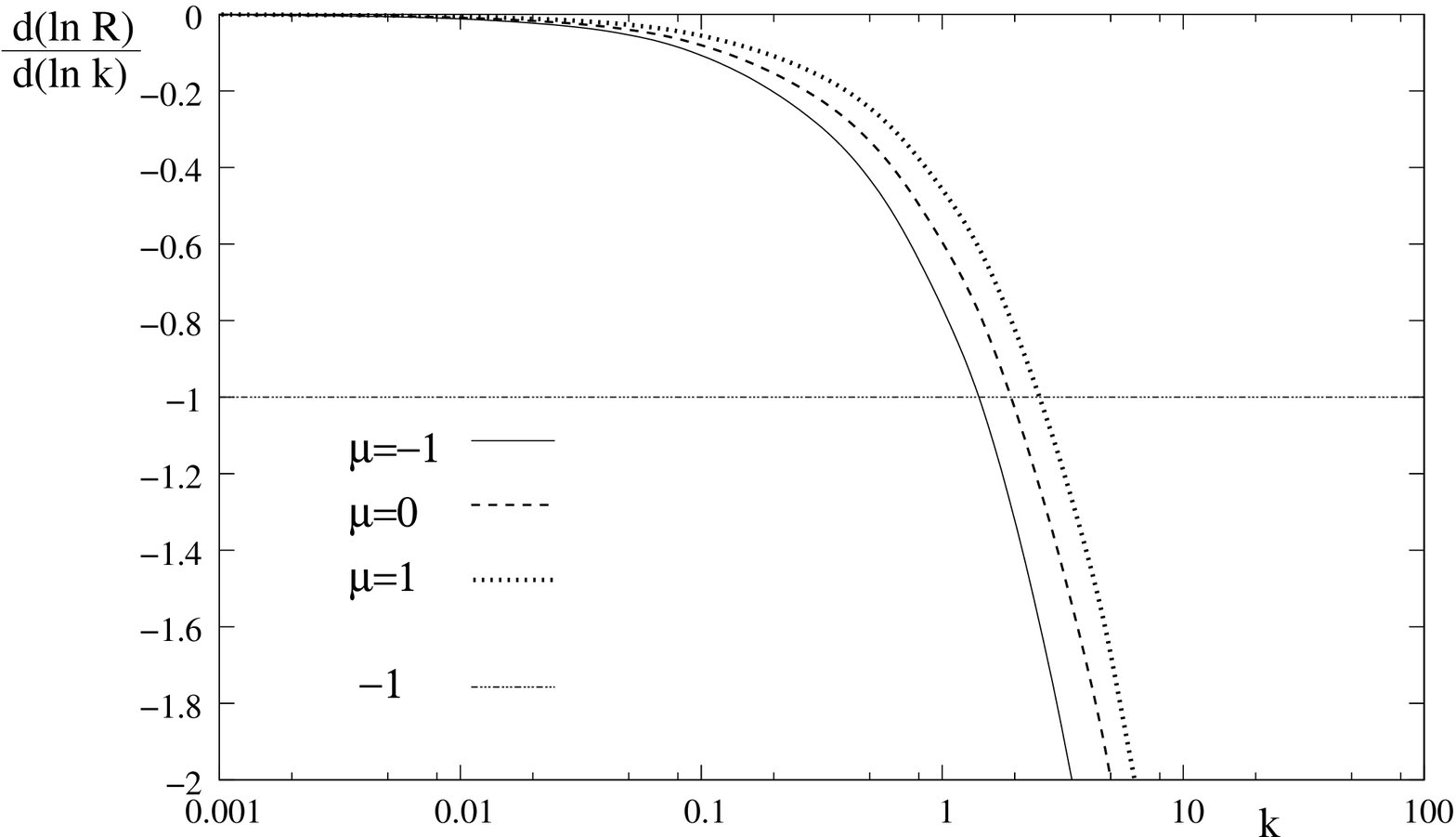}
	\caption[Calculating the cut-off momentum downstream. ($u_-=.5$)]
{Plotted along the $x$-axis we have the logarithm of momentum $k_+$ while 
along the $y$-axis we have $d(\ln R_+)/d(\ln k_+)$ for $u_-=.5$ and 
$R_+=\sum_i b^+_i(k_+)Q_i^+(k_+,\mu_+)$. Note the cut-off depends on $\mu_+$}
	\label{fig: cutoffu_5}
\end{figure}

\begin{figure}
 \centering
	\includegraphics[width=.9\columnwidth]{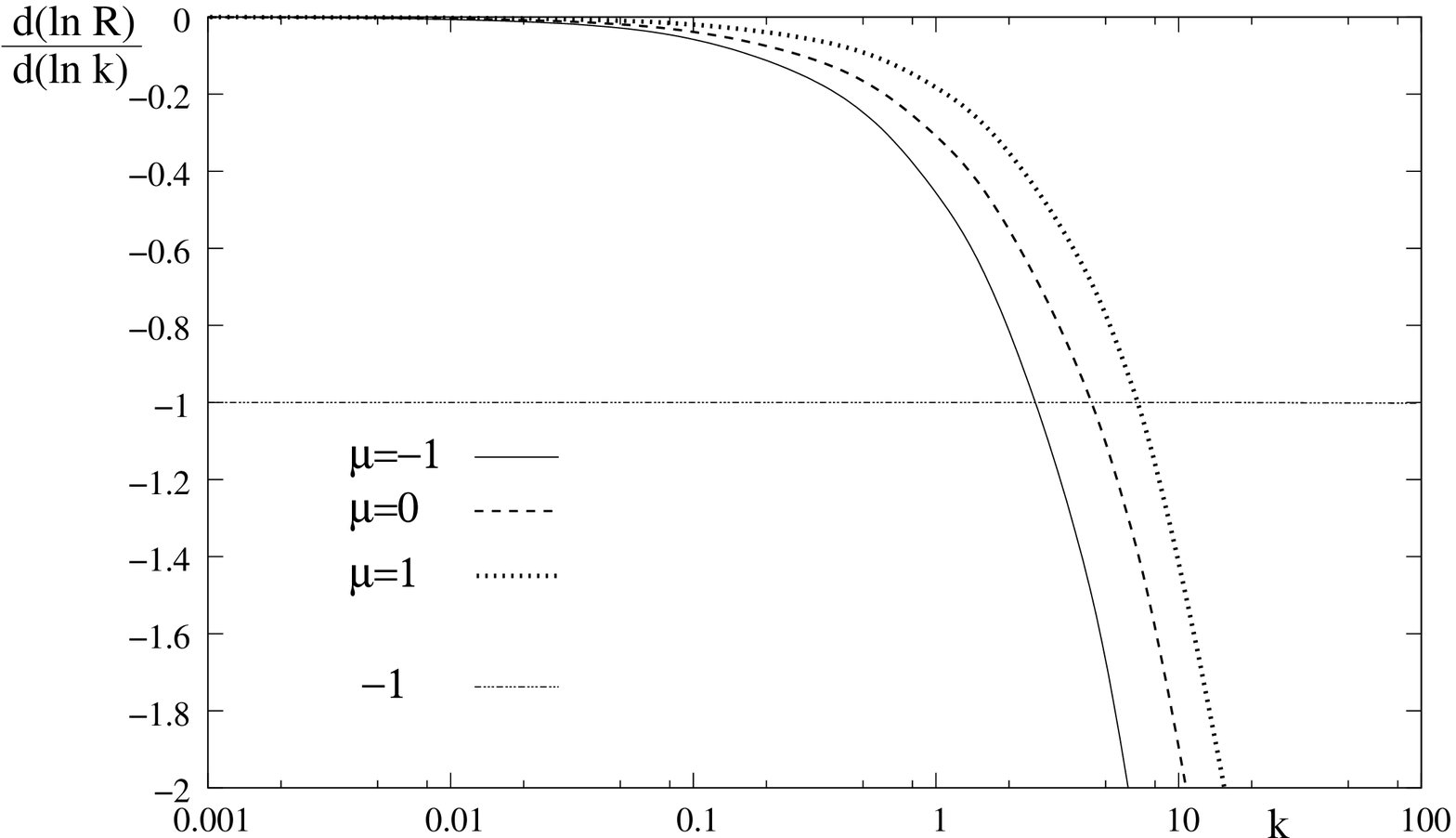}
	\caption[Calculating the cut-off momentum downstream. ($u_-=.7$)]
{Plotted along the $x$-axis we have the logarithm of momentum $k_+$ while 
along the $y$-axis we have $d(\ln R_+)/d(\ln k_+)$ for $u_-=.7$ and 
$R_+=\sum_i b^+_i(k_+)Q_i^+(k_+,\mu_+)$. Note the cut-off depends on $\mu_+$}
	\label{fig: cutoffu_7}
\end{figure}

Guided by the discussion for the nonrelativistic case, we use the expansion
\begin{align}
{\hat W}^{\pm} = \sum_{i}b^{\pm}_{i}k_\pm^{-s+3}Q^{\pm}_{i}(k_\pm,\mu_\pm)
\end{align}
so that the matching condition for the Laplace transformed spectrum at the shock reduces to 
\begin{align}
\Gamma_{\rm rel}^{s}(1+u_{\rm rel}\mu_-)^{s}\sum_{i}b_i^-(k_-)Q_{i}^-(k_-,\mu_-)=\nonumber\\
 \sum_{i}b_i^+(k_+)Q^{+}_{i}(k_+,\mu_+).
\end{align}
In order to solve for the particle spectrum, we multiply by
$(u_{+}+\mu_{+}){Q}^{+}_{j}(k_+,\mu_+) \; j \ge 0$ and 
integrate over $\mu_{+}$. Then for a fixed $k_-$ we have {
\begin{align}
\sum_{i}b_i^-(k_-)\int_{-1}^{1}&(1+u_{rel}\mu_-)^{s}Q_{i}^-(k_-,\mu_-)\times\nonumber\\&(u_{+}+\mu_{+}){Q}^{+}_{j}(k_+,\mu_+)d\mu_{+} = 0.
\end{align}}
Defining a matrix ${\bf S}$ with elements
\begin{align}
S^-_{i,j} =\int_{-1}^{1}(1+u_{\rm rel}\mu_-)^{s}Q_{i}^-(k_-,\mu_-)
(u_{+}+\mu_{+}){Q}^{+}_{j}(k_+,\mu_+)d\mu_{+}
\end{align}
we need to find the spectral index $s$, such that $\det {\bf S} = 0$. The Laplace inversion is 
then carried out numerically (see Appendix for details). 
As motivated by the nonrelativistic case, we define the cut-off 
to be the point at which 
\begin{align}
 \left.\diff{(\ln R)}{(\ln k)}\right|_{p_{\rm cut}} = -1
\end{align}
where $R=\sum b_iQ_i$. Figures \ref{fig: cutoffu_3}, \ref{fig: cutoffu_5} and 
\ref{fig: cutoffu_7} plot  $d(\ln R)/d(\ln k)$ at the shock against 
$k$ as measured downstream. The results are summarised in table 
\ref{table: cut_offs}.

\begin{table}
 \centering
\begin{tabular}{|l||l|l|l|}
\hline
 $u_-$ & .3 & .5 & .7 \\
\hline
 $u_+$ & .076 & .129 & .189 \\
\hline
$\Gamma_{\rm rel}$ & 1.027 & 1.089 & 1.23 \\
\hline
\hline
 $p^\ast$ & .404 & 1.143 & 2.26 \\ 
 Non-Rel $p_{\rm cut}$ & .621 & 1.79 & 3.719 \\
\hline
 $p_{\rm cut}(\mu_+=-1.0)$ & .541 & 1.42 & 2.566 \\ 
 $p_{\rm cut}(\mu_+=-0.5)$ & .595 & 1.71 & 3.612 \\ 
 $p_{\rm cut}(\mu_+=0)$ & .64 & 1.929 & 4.375 \\ 
 $p_{\rm cut}(\mu_+=0.5)$ & .682 & 2.138 & 5.105\\
 $p_{\rm cut}(\mu_+=1)$ & .741 & 2.533 & 6.773\\
\hline
\end{tabular}
\caption[Summary of Cut-Off Momenta]{Summary of Cut-Off Momenta}
\label{table: cut_offs}
\end{table}

\begin{figure}
 \centering
	\includegraphics[width=.9\columnwidth]{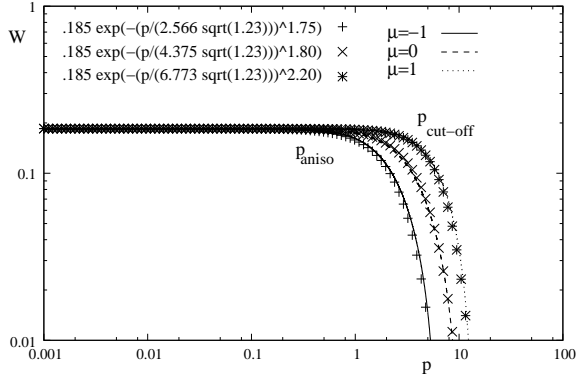}
	\caption[Distribution function at the shock, measured downstream. 
($u_-=.7$)]{The downstream function evaluated at the shock for a shock speed 
of .7. Along the $x$-axis we have plotted the logarithm of momentum $p_+$ 
while along the $y$-axis we have the logarithm of $W=p^4f$. The 
lines are data while the points are the best fit described in the text.}
	\label{fig: distribu_7shock}
\end{figure}

Figures \ref{fig: cutoffu_3}, \ref{fig: cutoffu_5} and \ref{fig: cutoffu_7} 
show how the cut-off momentum becomes increasing anisotropic as 
the shock speed increases.
\begin{figure}
 \centering
	\includegraphics[width=.9\columnwidth]{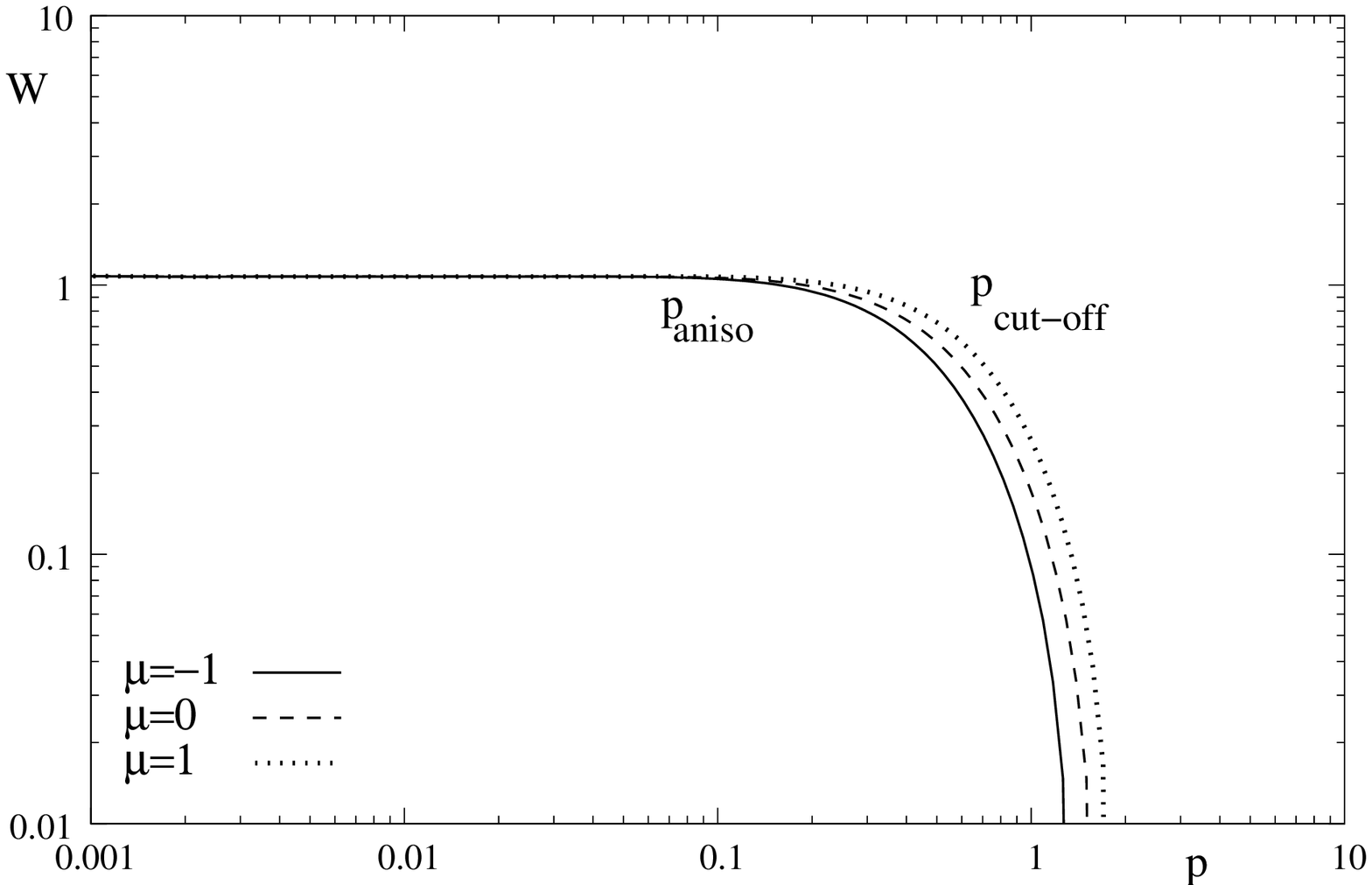}
	\caption[Distribution function at the shock, measured downstream. 
($u_-=.3$)]{The downstream function evaluated at the shock for a shock 
speed of .3. Along the $x$-axis we have plotted the logarithm of momentum 
$p_+$ while along the $y$-axis we have the logarithm of $W=p^4f$.}
	\label{fig: distribu_3shock}
\end{figure}
\begin{figure}
 \centering
	\includegraphics[width=.9\columnwidth]{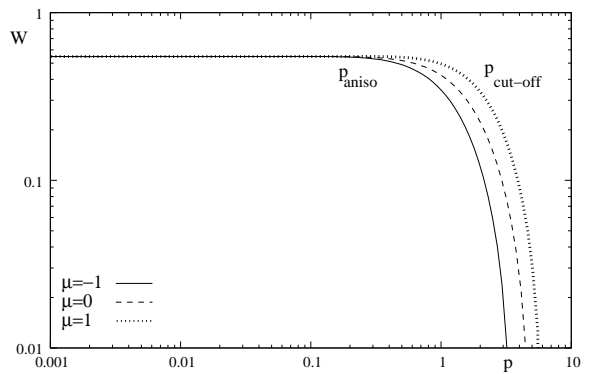}
	\caption[Distribution function at the shock, measured downstream. 
($u_-=.5$)]{The downstream function evaluated at the shock for a shock 
speed of .5. Along the $x$-axis we have plotted the logarithm of momentum 
$p_+$ while along the $y$-axis we have the logarithm of $W=p^4f$.}
	\label{fig: distribu_5shock}
\end{figure}
The distribution can be fitted approximately by 
\begin{align}
f\approx p^{-s} \exp\left(-\left(\frac{p}{\sqrt{\Gamma_{\rm rel}}
p_{\rm cut}(\mu_+)}\right)^\beta\right)
\end{align}
where $\beta$ is typically 2. However it is difficult justify the use of 
the factor $\sqrt{\Gamma_{\rm rel}}$ in general as our results are only for 
mildly relativistic shocks. This fit justifies our definition of $p_{\rm cut}$ 
instead of using the equilibrium momentum $p^\ast$. 
Figure \ref{fig: distribu_7shock} illustrates this approximation for a .7c 
shock. $\beta$ seems to be pitch angle dependent varying between 1.75 and 2.2, 
but typically 2. In fact for the .3c and .5c shock cases $\beta$ showed much 
less variation about 2. 
For the shock speeds we have chosen, with the Juttner-Synge equation of state, 
the spectral indices in the absence of losses are close to $4$.

Figures~\ref{fig: distribu_3shock}, ~\ref{fig: distribu_5shock} and ~\ref{fig: distribu_7shock} illustrated a feature that was not present in the non-relativistic case. The pitch angle dependence of the cut-off momentum leads to a difference in the isotropy levels between particles above and below some critical momentum $p_{\rm aniso}$. Indeed there is a clear pattern of greater levels of anisotropy at high energies as the shock speed increases, despite the fact that the results presented here are only for mildly relativistic shocks. 

\subsection{The Spatially Integrated Distribution}
While the method we follow in this paper finds the upstream particle 
distribution directly, it is easy to find the downstream distribution 
by using the matching condition, as discussed in the previous section. The 
downstream distribution is, in many respects, more important physically as 
it will be responsible form most of the spatial integrated emission. As it 
can be difficult to spatially resolve observational data from non-thermal 
emitters, we must consider the emission from an extended region of space. 
Our eigenfunction expansion allows us to do this quite easily. The spatially 
averaged distribution from a downstream region $[z_0,z_1]$ in terms of Laplace 
variables is 
\begin{align}
R_{[z_0,z_1]}(k_+,\mu_+)=\sum_{i\le0}a_i(k)\left(\exp\left(\frac{\Lambda^+_i(k_+)z_1}{\Gamma_+}\right)
\nonumber\right. \\ -\left.\exp\left(\frac{\Lambda^+_i(k_+)z_0}{\Gamma_+}\right)\right)\frac{\Gamma_+Q^+_i(\mu_+,k_+)}{\Lambda^+_i(k_+)}.
\end{align}
In the case of a source which is completely spatially unresolved this reduces to
\begin{align}
R_{[0,\infty]}(k_+,\mu_+)=-\Gamma_+\sum_{i\le0}a_i(k)\frac{Q^+_i(\mu_+,k_+)}{\Lambda^+_i(k_+)}.
\end{align}
Of course the optical depth of the emitting region will also have an effect 
on the spectrum of unresolved sources by reducing $z_1$.

\begin{figure}
 \centering
	\includegraphics[width=.9\columnwidth]{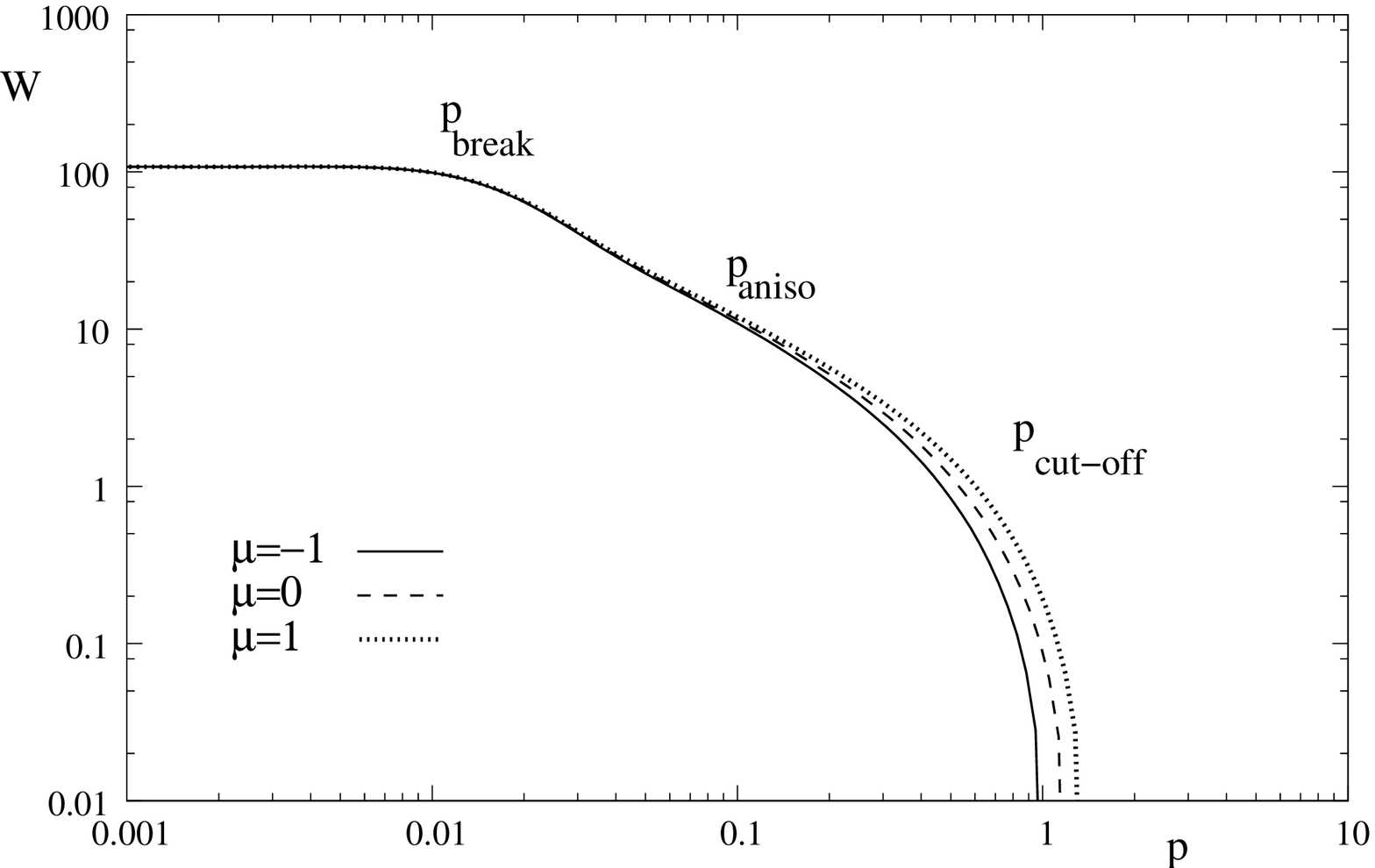}
	\caption[Integrated distribution function behind a shock with $u_− = .3$ .]
{The downstream function integrated for a shock speed of .3 between $z'=0$ 
and $z'=100$ where $z'=Dz/\Gamma_+$. Along the $x$-axis we have plotted the 
logarithm of momentum $p_+$ while along the $y$-axis we have the logarithm 
of $W=p^4f$.}
	\label{fig: distribu_3INT}
\end{figure}

\begin{figure}
 \centering
	\includegraphics[width=.9\columnwidth]{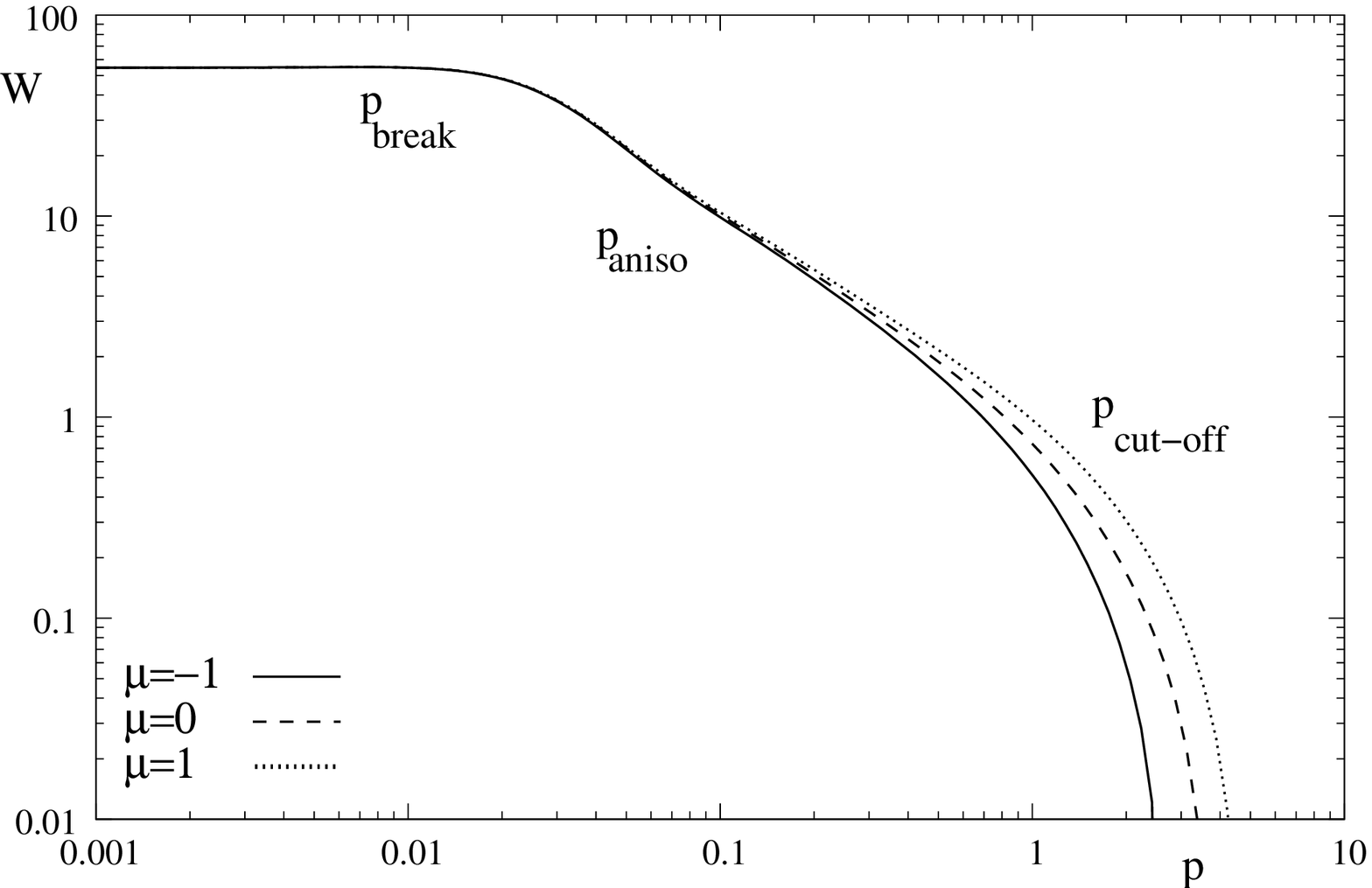}
	\caption[Integrated distribution function behind a shock with $u_− = .5$]
{The downstream function integrated for a shock speed of .5 between $z'=0$ 
and $z'=100$ where $z'=Dz/\Gamma_+$. Along the $x$-axis we have plotted the 
logarithm of momentum $p_+$ while along the $y$-axis we have the logarithm 
of $W=p^4f$.}
	\label{fig: distribu_5INT}
\end{figure}

\begin{figure}
 \centering
	\includegraphics[width=.9\columnwidth]{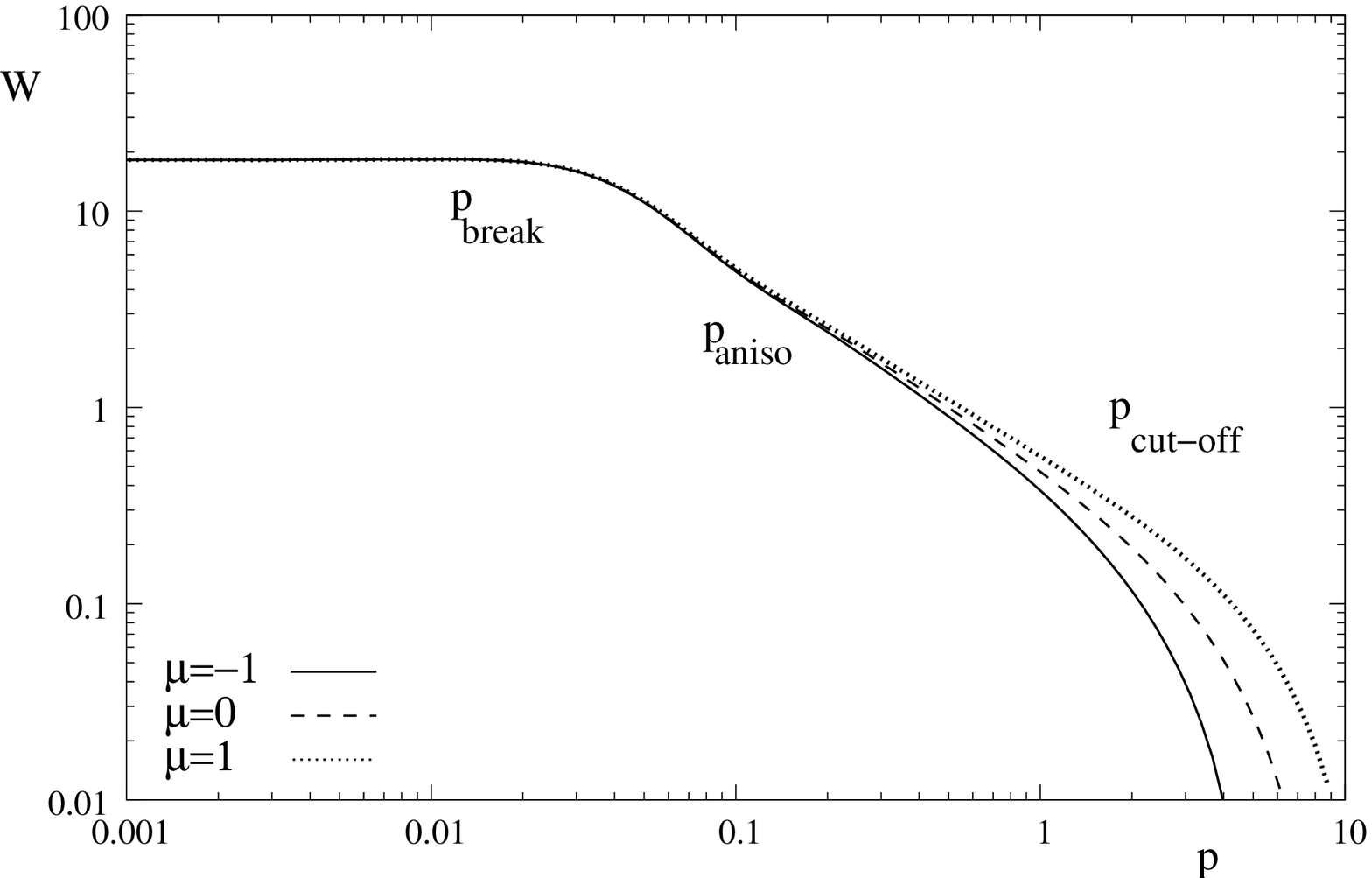}
	\caption[Integrated distribution function behind a shock with $u_− = .7$]
{The downstream function integrated for a shock speed of .7 between $z'=0$ 
and $z'=100$ where $z'=Dz/\Gamma_+$. Along the $x$-axis we have plotted the 
logarithm of momentum $p_+$ while along the $y$-axis we have the logarithm 
of $W=p^4f$.}
	\label{fig: distribu_7INT}
\end{figure}

\begin{figure}
 \centering
	\includegraphics[width=.9\columnwidth]{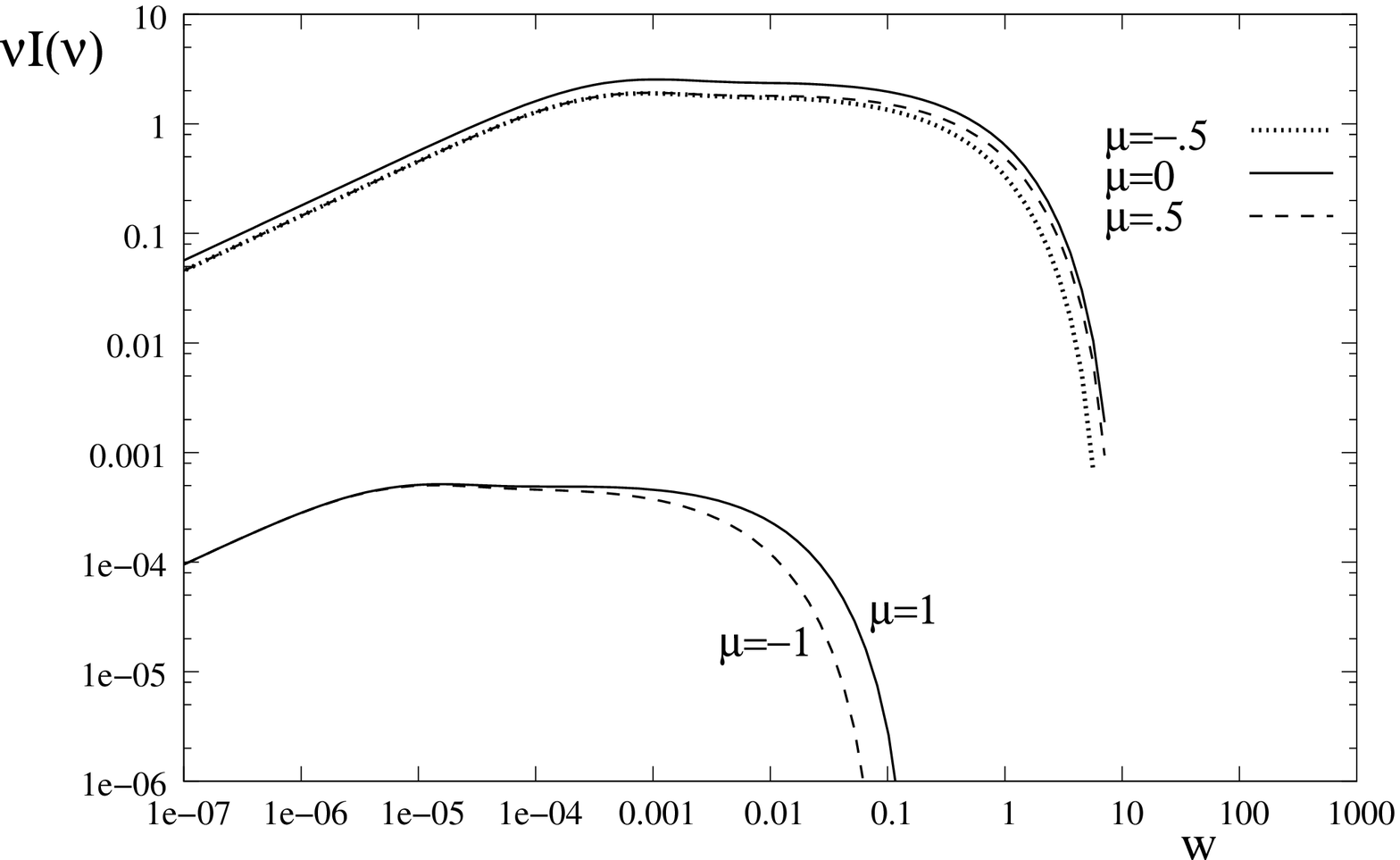}
	\caption[Spatially integrated synchrotron emission from a shock with 
$u_-=.3$]{Synchrotron emission from the particle distribution shown in 
\ref{fig: distribu_3INT} measured in the downstream medium. In plotting our 
$\mu=\pm1$ we used $\mu=\pm.9999$ as there is no emission from an ordered 
field along $\mu=\pm1$.}
	\label{fig: nuInu_u3INT}
\end{figure}

\begin{figure}
 \centering
	\includegraphics[width=.9\columnwidth]{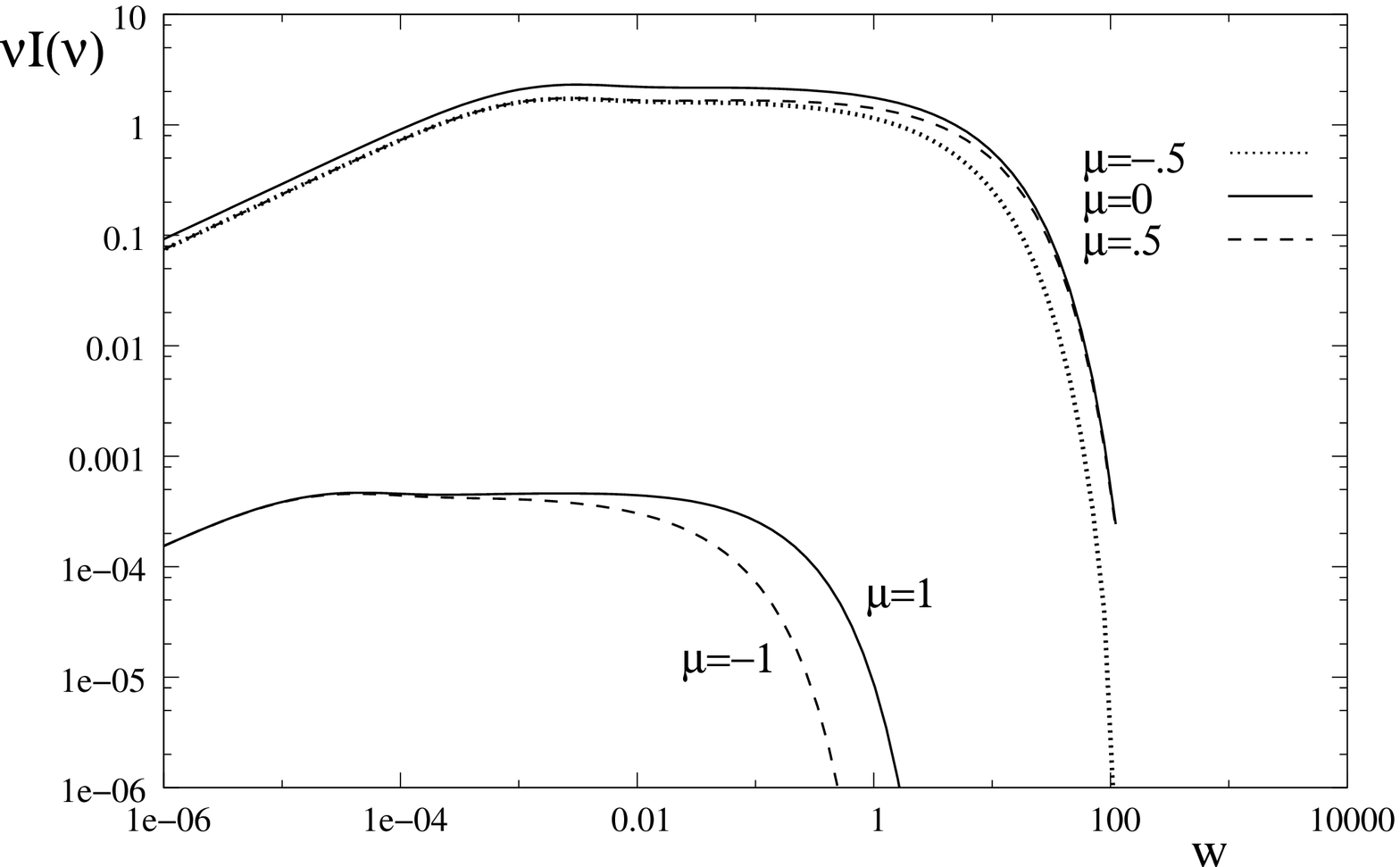}
	\caption[Spatially integrated synchrotron emission from a shock 
with $u_-=.5$]{Synchrotron emission from the particle distribution 
shown in \ref{fig: distribu_5INT} measured in the downstream medium. 
In plotting our $\mu=\pm1$ we used $\mu=\pm.9999$ as there is no 
emission from an ordered field along $\mu=\pm1$.}
	\label{fig: nuInu_u5INT}
\end{figure}

\begin{figure}
 \centering
	\includegraphics[width=.9\columnwidth]{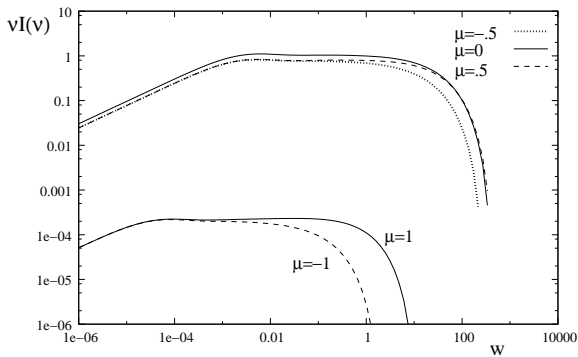}
	\caption[Spatially integrated synchrotron emission from a shock 
with $u_-=.7$]{Synchrotron emission from the particle distribution shown 
in \ref{fig: distribu_7INT} measured in the downstream medium. In plotting 
our $\mu=\pm1$ we used $\mu=\pm.9999$ as there is no emission 
from an ordered field along $\mu=\pm1$.}
	\label{fig: nuInu_u7INT}
\end{figure}

Using the same numerical Laplace inversion as in the non-relativistic case we have calculated the distribution 
functions and synchrotron emission. Figures~\ref{fig: distribu_3INT},~\ref{fig: distribu_5INT} and~\ref{fig: distribu_7INT} 
show the spatially integrated distribution functions for a finite emission region. Now there are there features: a momentum 
break, $p_{\rm b}$, due to spatial effect; an anisotropic break, $p_{\rm aniso}$, due to relativistic effects; and a cut-off, 
$p_{\rm cut}$, due to energy losses. Given that the magnetic field is constant throughout this region it is trivial to 
produce the associated synchrotron emission plots of figures~\ref{fig: nuInu_u3INT},~\ref{fig: nuInu_u5INT} 
and~\ref{fig: nuInu_u7INT}. It should be noted that in the emission plots $I_\nu$ is measured in the downstream 
frame, but since $I_\nu/\nu^3$ is a Lorentz invariant the transformation is trivial. The synchrotron emission also 
includes the same three features we observed in the particle distribution; namely a break frequency beyond which the 
effect of synchrotron cooling becomes important, a frequency at which pitch-angle or anisotropic effects play a role
and an upper cut-off beyond which there is virtually no emission.

\section{Discussion}
Particle acceleration and self-consistent synchrotron radiation have been considered previously by 
\citet{JKFRAM1998} using a zonal model. They were successful in explaining the radio to X-ray spectrum of 
Mkn 501. However such zonal models typically depend on isotropic particle distributions. We have shown, 
however, that for particles near the high energy cut-off this is not true even for mildly relativistic 
flows. The computational resources available restricted our results to be below .7c. However even for the mildly 
relativistic shock velocities we see a clear pattern of high energy anisotropy emerging
resulting in synchrotron emission which is also anisotropic. 
This could be extremely important in the modelling of the inverse Compton hump in $\gamma$-rays observed in 
TeV Blazars \citep{HESS2006AA455}.
As a second implication of the particle anisotropy, in the presence of losses,
the idealised situation, of a two sided strongly polarised 
identical jet system can be considered. Each jet contains only forward external shocks, and the jet which is directed 
towards the observer is inclined at an angle $\theta=\cos^{-1} (-\mu)$ to the line of sight (magnetic 
field direction same as that of shock).
Then we will observe the emission from particles in the jet directed towards 
us which have pitch angle $\mu$ and from particles in the jet directed 
away from us which have pitch angle $-\mu$. While at low energies the only difference between the observed emission 
of the two jets will be as a result of the effects of beaming, at energies near the synchrotron cut-off the details 
of the acceleration mechanism will amplify this difference, depending on viewing angle.

Although the work in this paper is limited to an idealised form of diffusion, and mildly relativistic shocks, 
it illustrates previous unexamined features which could be important in the modelling of relativistic, 
$\gamma$-ray sources such as microquasars, blazars and GRBs. We have parameterised the exponential shape 
of the distribution cut-off and identified new pitch angle dependent features between break and cut-off 
frequencies. Further work is needed to examine both momentum dependent scattering and high Lorentz 
factor flows.

\section*{Acknowledgments}

Paul Dempsey would like to thank the Irish Research Council for Science, 
Engineering and Technology for their financial support. 
He would also like to thank Cosmogrid for access to their computational facilities.
We are grateful for discussions with Felix Aharonian. Peter Duffy would like to thank the
Dublin Institute for Advanced Studies for their hospitality during the completion 
of this work. We would like to thank the referee for comments that improved the quality of this paper.

\bsp
\appendix
\section{Inverse Laplace Transforms}

While \citet{HeavensMeisenheimer87} invert the Laplace transform analytically 
for particular cases here we use  numerical methods as we will need 
to when dealing with relativistic flows. 

Formally the inverse Laplace transform is the Bromwich integral, which is a 
complex integral given by:
\begin{align}
 f(t) = \mathcal{L}^{-1}\left[F(s)\right]=\frac{1}{2\pi i}\int_{\gamma -i\infty}^{\gamma +i\infty} e^{st} F(s)\;ds
\end{align}
where $\gamma$ is to the right of every singularity of $F(s)$. If the 
singularity of $F(s)$ all ly in the left half of the complex plane $\gamma$ 
can be set to $0$ and this reduces to the inverse Fourier transform, which is 
easy to do. However for complicated or numerical Laplace functions the 
Bromwich integral is extremely difficult to solve. The four main numerical 
inversion techniques are Fourier Series Expansion, Talbot's method, Weeks 
method and methods based on the Post-Widder formula. However some of these 
methods converge rather slowly and a lot of work has gone into creating 
acceleration methods. Numerical Laplace inversion is a area of active 
research and the choice of inversion technique is as much an art as a 
science at the moment. In this paper 
the Post-Widder based method was
chosen and only these methods shall by described below.
Let $F(s)$ be the Laplace transform of $f(t)$ then \citet{Widder1932} showed 
that $f_n(t)\rightarrow f(t)$ where 
\begin{align}
f_n(t) = \frac{(-1)^n }{n!} \left(\frac{n+1}{t}\right)^{n+1} \hat{f}^{(n)}((n+1)/t).
\end{align}
The advantage of this method in our case in that we see that the Laplace 
transform of the solution times the Laplace coordinate is the zeroth order 
approximation to the actual solution.
\begin{align}
&W_0(y) = \left(\frac{1}{y}\right) \hat{W}(1/y)\nonumber
\Rightarrow W_0(p)= p\hat{W}(p).
\end{align}
When dealing with numerical results however it is easier to use the 
Gaver-Stehfest algorithm \citep{AV2004}. It is an algorithm based on the 
Post-Widder method with the Gaver approximants, $\{f_n(t) : n\ge0\}$, defined as
\begin{align}
 f_n(t) \equiv \frac{(n+1) \ln(2)}{t} \binom{2(n+1)}{n+1}\times\nonumber\\\sum_{k=0}^{n+1}\binom{n+1}{k}\hat{f}\left((n+1+k)\ln(2)/t\right).
\end{align}
However the convergence for both these methods is slow. A test of methods for 
accelerating this convergence can be found in \citet{Valko2004} and two are 
found to be quite good: the non-linear Wynn's Rho Algorithm and the linear 
Salzer summation. Again a choice has to be made and here we present only 
Salzer summation: $f(t,M)\rightarrow f(t)$ where
\begin{align}
f(t,M)=\sum_{k=1}^M W_k f_{k-1}(t) 
\end{align}
and 
\begin{align}
 W_k = (-1)^{k+M} \frac{k^M}{M!}\binom{M}{k}.
\end{align}
The Post-Widder method based on differentiation was implemented in Maple with 
the Salzer acceleration. It was used to produce the results in the 
non-relativistic limit as we have an analytic form of the Laplace function to work with.
The Salzer accelerated Gaver-Stehfest algorithm was implemented in C/C++ code 
for use with the numerical output from the relativistic approach discussed above. 

{
\section{Deriving the eigensystem differential equations}
The solutions, $Q_i$, to equation \ref{SLeqnQ}
\begin{align}
\pdiff{}{\mu}\left(D(\mu)(1-\mu^2)\pdiff{Q_i }{\mu}\right)
-x\lambda g(\mu)Q_i = \Lambda_i(u+\mu)Q_i
\end{align}
for real $x$, are orthogonal, with weight $u+\mu$ and have real, distinct eigenvalues $\Lambda_i$ \citep{Boas83}.
Taking the derivative of this equation with respect to $x$ gives
\begin{align}
\pdiff{}{\mu}\left(D(\mu)(1-\mu^2)\pdiff{}{\mu}\pdiff{Q_i}{x}\right)
-x\lambda g(\mu)\pdiff{Q_i}{x} \nonumber\\= \Lambda_i(u+\mu)\pdiff{Q_i}{x} + \left(\diff{\Lambda_i}{x}(u+\mu)+\lambda g(\mu)\right)Q_i.
\end{align}
Since solution to Sturm Liouville equations form an orthogonal basis, we can write
\begin{align}
 \pdiff{Q_i}{x}=\sum_m q_mQ_m
\end{align}
which gives 
\begin{align}
\sum_m q_m\pdiff{}{\mu}\left(D(\mu)(1-\mu^2)\pdiff{}{\mu}Q_m\right)
-\sum_m q_m x\lambda g(\mu) Q_m \nonumber\\= 
\sum_m q_m \Lambda_m(u+\mu)Q_m\nonumber\\=
\sum_m q_m\Lambda_i(u+\mu) Q_m + \left(\diff{\Lambda_i}{x}(u+\mu)+\lambda g(\mu)\right)Q_i.
\end{align}
Multiplying by $Q_j^*$ and integrating over $\mu$ gives
\begin{align}
\sum_m q_m (\Lambda_m-\Lambda_i)\langle Q_j, Q_m \rangle \nonumber\\= \diff{\Lambda_i}{x}\langle Q_j, Q_i \rangle + \lambda \int_{-1}^1 g(\mu)Q_iQ_j^*\;d\mu .
\end{align}
Taking $j=i$ implies equation \ref{Evaluederiv} and $j\ne i$ implies equation \ref{Efunctderiv}.
}

\label{lastpage}

\end{document}